\documentclass{article}
\usepackage{arxiv}

\usepackage{amsmath, amssymb, bm, graphicx, url}
\usepackage[authoryear, round]{natbib}
\usepackage{hyperref}
\providecommand{\dodoi}[1]{\href{https://doi.org/#1}{doi:#1}}

\usepackage{pdfpages}

\newcommand{\dd}{\mathrm{d}}
\newcommand{\ii}{\mathrm{i}}

\newcommand{\TT}{\mathrm{T}}

\title{Probabilistic wavefield decomposition from array covariance}
\author{Jack-Andrew Smith \\
  School of GeoSciences, University of Edinburgh, Edinburgh, UK \\
  \texttt{jack-andrew.smith@ed.ac.uk}
  \And
  Claudia Elijas-Parra \\
  School of GeoSciences, University of Edinburgh, Edinburgh, UK}

\begin{document}

%% Version in square brackets will appear at top right of all pages except title page:
%\title[Probabilistic wavefield decomposition]{Probabilistic wavefield decomposition from array covariance}

\maketitle

\begin{abstract}

Inter-sensor coherence loss carries information about the wavefield that conventional array processing discards. The array-covariance matrix is modelled using a mixture of energy distributions over slowness with analytical characteristic functions, e.g., skew-normal and angular-harmonic distributions, and inverted using variational Bayesian inference. The posterior supports uncertainty quantification, experimental design, model comparison and hypothesis testing. For a synthetic hydroacoustic T-phase, information is increased by an additional sensor and decreased by noise; the data favour a distributed source over a point source. For an observed thermospheric infrasound return, energy is skewed back towards the source. The method is implemented in the software Topophone2.

\end{abstract}

\maketitle

%  End of title page for JASA-EL  --  --  --  --  --  --  --  --  --  --  --  %

\section{Introduction}

A principal objective of array processing is to estimate the apparent velocity and direction of arrival of incoming signals, represented in slowness space. Many array-processing methods, such as conventional and adaptive beamforming \citep{Bartlett_1948, Capon_1969}, assume a far-field point source. Many detectors with calibrated statistics, such as the F-detector \citep{Blandford_1974}, further assume that noise is incoherent between sensors. The former assumption is invalid for spatially distributed primary sources, and for multipathing or secondary sources associated with scattering through heterogeneous media \citep{Valaee_et-al_1995}; the latter is invalid when ambient environmental noise propagates coherently across the array \citep{Selby_2008}.

Wavefields distributed over slowness cause inter-sensor coherence loss, which is often treated as a nuisance in array processing \citep{Gibbons_et-al_2008}. Given an analytical model for this coherence loss, however, the cross-spectral densities between sensors in an array -- the array covariance -- can instead be inverted for a parameterised distribution of energy over slowness \citep{Valaee_et-al_1995}. Such models exist for degenerate (plane waves), bounded uniform \citep{Mack_Flinn_1971}, normal \citep{Gossard_1969}, normal-mixture \citep{Wu_et-al_2019} and angular-harmonic \citep{Cox_1973} distributions. Models for asymmetric distributions are necessary if the primary or secondary sources are asymmetrically distributed in space, or if the wavefield is scattered within a moving medium.

Most existing covariance- and delay-based inversions provide a single best-fitting model, with uncertainties that are at best local or asymptotic approximations, often conditional on fixed nuisance parameters \citep[e.g.,][]{Trump_Ottersten_1996,Szuberla_Olson_2003,Green_2015,Poste_et-al_2023}. Yet the inverse problem is nonlinear, non-unique and often ill-conditioned, requiring \textit{ad hoc} regularisation to stabilise deterministic solutions, which biases both the estimate and its uncertainty \citep{Galetti_et-al_2015}. These uncertainties matter: backazimuth uncertainty propagates directly into source-location uncertainty \citep{Modrak_et-al_2010}, which underpins monitoring decisions \citep{Anderson_et-al_2007,Williams_et-al_2026}.

In this letter, we generalise the forward model for the array covariance to mixtures of incoherently distributed energy densities with analytical characteristic functions, e.g., skew-normal signals and angular-harmonic environmental noise. We pose the inverse problem probabilistically and compute the posterior using variational inference with normalising flows. Synthetic hydroacoustic T-phase and observed thermospheric infrasound return examples demonstrate how the posterior supports uncertainty quantification, experimental design, model comparison and hypothesis testing. The method is implemented in the open-source Python package Topophone2 \citep{Smith_Elijas-Parra_2026}.

\section{Probabilistic wavefield decomposition}

\subsection{\label{sec:forward}Forward problem}

Consider an array of sensors recording a scalar component of a temporally stationary wavefield, such as acoustic pressure, with sensor~$i$ located at $\bm{x}_i$. The wavefield is represented as a superposition of components $k$, each with total energy $E_k$ incoherently distributed over slowness $\bm{s}$ according to the normalised density $f_k(\bm{s} \mid \theta_k)$ parameterised by $\theta_k$. In practice, a parsimonious parameterisation of the wavefield makes inference of its properties tractable from a finite set of indirect measurements \citep{Sober_2015}, but must be balanced against fidelity to an often complex wavefield \citep[e.g.,][]{Xenaki_et-al_2016}. At angular frequency $\omega$, the recording by sensor $i$ is
\begin{equation}
    \label{eq:u}
    u_i(\omega) = \sum\nolimits_k \int \sqrt{E_k f_k(\bm{s} \mid \theta_k)} \, \exp\{\ii[\omega \bm{x}_i \cdot \bm{s} + \psi_k(\bm{s})]\} \, \dd\bm{s} \, + \, \sqrt{E_\varepsilon^{(i)}} \, \exp\{\ii \psi_\varepsilon^{(i)}\},
\end{equation}
where $\ii$ is the imaginary unit and $\psi_k(\bm{s})$ is the random phase of the plane wave with slowness $\bm{s}$ in component $k$. The self-noise component $\varepsilon$ at sensor $i$ has random phase $\psi_\varepsilon^{(i)}$ and energy $E_\varepsilon^{(i)}$, the latter assumed equal between all sensors in the following analysis. For brevity, the notation denoting frequency dependence of all quantities is suppressed. The cross-spectral density between sensors $i$ and $j$ is $\Gamma_{ij}(\omega) = \mathbb{E}[u_i(\omega) \, u_j^*(\omega)]$, where $\mathbb{E}[\cdot]$ denotes expectation and $*$ denotes complex conjugation. In an incoherently distributed wavefield, plane waves with different slownesses, or belonging to different components, are mutually uncorrelated, such that $\mathbb{E}\{\exp[\ii(\psi_k(\bm{s}) - \psi_{k'}(\bm{s}'))]\} = \delta_{kk'} \, \delta(\bm{s} - \bm{s}')$, where $\delta_{kk'}$ is the Kronecker delta and $\delta(\cdot)$ is the Dirac delta function \citep{Valaee_et-al_1995}. All cross-terms therefore vanish under the expectation, and with separation $\bm{r}_{ij} = \bm{x}_i - \bm{x}_j$ the cross-spectral density reduces to
\begin{equation}
    \label{eq:Gamma}
    \Gamma_{ij}(\omega \mid \Theta) = \sum\nolimits_k E_k \int f_k(\bm{s} \mid \theta_k) \exp(\ii \omega \bm{r}_{ij} \cdot \bm{s}) \, \dd\bm{s} \, + \, E_\varepsilon \delta_{ij} = \sum\nolimits_k E_k \, \gamma_k(\omega \bm{r}_{ij} \mid \theta_k) + E_\varepsilon \gamma_\varepsilon,
\end{equation}
where $\gamma_k$, the coherency of component $k$, is the characteristic function of $f_k$ evaluated at $\omega \bm{r}_{ij}$ ($\gamma_\varepsilon = \delta_{ij}$ for the self-noise component). The full parameter set over all components, including the energies, is $\Theta$. We treat $\Theta$ as constant within each narrow frequency band analysed in Sec.~\ref{sec:examples}; broader bands could be accommodated by parameterising $\Theta$ as a function of frequency. The cross-spectral densities of all sensor pairs form the array-covariance matrix $\bm{\Gamma}$, with elements $\Gamma_{ij}$. In practice, the array-covariance matrix is estimated by the sample-covariance matrix $\hat{\bm{\Gamma}}$, an arithmetic average over $N$ independent realisations. These may be obtained from separate time windows or from orthogonal Slepian tapers \citep{Thomson_1982}.

Closed-form characteristic functions are known for many parameterised distributions $f_k$ \citep{Oberhettinger_1973}. For example, the skew-normal distribution with location vector $\bm{\xi}$, scale matrix $\bm{\Omega}$ and shape vector $\bm{\alpha}$ is
\begin{equation}
    \label{eq:SN}
    f_{\mathrm{SN}}(\bm{s} \mid \bm{\xi}, \bm{\Omega}, \bm{\alpha}) = 2 \, \phi(\bm{s} \mid \bm{\xi}, \bm{\Omega}) \, \Phi[\bm{\alpha}^\TT \bm{W}^{-1} (\bm{s} - \bm{\xi})],
\end{equation}
where $\phi$ is the multivariate normal density, $\Phi$ is the standard normal cumulative distribution function, and $\bm{W} = \operatorname{diag}(\bm{\sigma})$ contains the beamwidths $\bm{\sigma}$, such that $\bm{\Omega} = \bm{W} \bm{\Omega}_z \bm{W}$ with correlation matrix $\bm{\Omega}_z$. \citet{Kim_Genton_2011} give the corresponding characteristic function as
\begin{equation}
    \label{eq:SN_CF}
    \gamma_{\mathrm{SN}}(\omega \bm{r}_{ij} \mid \bm{\xi}, \bm{\Omega}, \bm{\alpha}) = \exp(\ii \omega \bm{r}_{ij}^\TT \bm{\xi} - \omega^2 \bm{r}_{ij}^\TT \bm{\Omega} \bm{r}_{ij} / 2) \, [1 + \ii \operatorname{erfi}(\omega \bm{d}^\TT \bm{W} \bm{r}_{ij} / \sqrt{2})],
\end{equation}
where $\bm{d} = \bm{\Omega}_z \bm{\alpha} / (1 + \bm{\alpha}^\TT \bm{\Omega}_z \bm{\alpha})^{1/2}$. The limit $\bm{\alpha} \to \bm{0}$ recovers the coherency of a normally distributed wavefield, and the limit $\bm{\sigma} \to \bm{0}$ recovers a single plane wave. Skewness introduces an additional phase, $\arg[1 + \ii \operatorname{erfi}(\cdot)]$, centring the cross-phase on the mean -- rather than the location or mode -- in the limit where the skew $\bm{W} \bm{d} \ll |\omega\bm{r}_{ij}|^{-1}$ [Sec.~S1, supplementary material]. In theory, the coherency of any energy-density distribution can be approximated using a weighted sum of normal characteristic functions. For interpretation, the location is quoted in polar coordinates, while the beamwidth and skew are quoted as radial and transverse components about $\bm{\xi}$; the transverse component is expressed as an angle by dividing by $|\bm{\xi}|$ under the small-angle approximation.

Wavefield components of interest are referred to as ``signal'' components, otherwise they are ``noise'' components. Often, in addition to sensor self-noise, the ambient environmental wavefield, which propagates coherently across the array, is considered noise \citep[e.g.,][]{Selby_2008}. The directional energy density of a 2D environmental wavefield travelling at apparent (horizontal) slowness $s$ can be expanded in angular harmonics,
\begin{equation}
    \label{eq:noise_2D}
    f_\text{noise}(\varphi \mid \bm{a}, \bm{b}) = (2\pi)^{-1} \sum\nolimits_{m=0}^\infty \epsilon_m [a_m \cos(m\varphi) + b_m \sin(m\varphi)],
\end{equation}
where $m$ is the cylindrical-harmonic order, $\bm{\epsilon} = (1, 2, 2, \dots)$, $a_0 = 1$ for normalisation and the propagation azimuth is $(90^\circ - \varphi)$, defined clockwise from north. The corresponding coherency is
\begin{equation}
    \label{eq:noise_CF_2D}
    \gamma_\text{2D}(\omega \bm{r}_{ij} \mid \bm{a}, \bm{b}, s) = \sum\nolimits_{m=0}^\infty \ii^m \epsilon_m [a_m \cos(m\varphi_{ij}) + b_m \sin(m\varphi_{ij})] \, J_m(\omega |\bm{r}_{ij}| s),
\end{equation}
where $(90^\circ - \varphi_{ij})$ is the azimuth of $\bm{r}_{ij}$, and $J_m$ is the $m$th-order cylindrical Bessel function of the first kind \citep{Cox_1973}. In practice, the harmonic expansion is truncated to represent low-order anisotropy in the directional energy density; for isotropic fields only the $m = 0$ term remains. Expressions for the 3D angular-harmonic field are given by \citet{Cox_1973}.

\subsection{\label{sec:inverse}Inverse problem}

We calculate the full posterior probability density function (pdf) of the parameters $\Theta$ in order to appraise the full information in the observed sample-covariance matrix $\hat{\bm{\Gamma}}$. The posterior pdf is given by Bayes' rule, $p(\Theta \mid \hat{\bm{\Gamma}}) = p(\hat{\bm{\Gamma}} \mid \Theta) \, p(\Theta) / p(\hat{\bm{\Gamma}})$, which updates the prior pdf $p(\Theta)$ with the information contained in the data. The evidence $p(\hat{\bm{\Gamma}})$ normalises the posterior, and the likelihood $p(\hat{\bm{\Gamma}} \mid \Theta)$ gives the probability of observing $\hat{\bm{\Gamma}}$ given $\Theta$. Assuming that (a) the $N$ realisations are independent, with $N$ at least the number of sensors, (b) the Fourier coefficients of the recording follow a zero-mean multivariate complex Gaussian distribution with circular symmetry, and (c) the array-covariance matrix is positive definite, \citet{Goodman_1963} showed that $N\hat{\bm{\Gamma}}$ follows a complex Wishart distribution,
\begin{equation}
    \label{eq:likelihood}
    p(\hat{\bm{\Gamma}} \mid \Theta) \propto |\bm{\Gamma}(\Theta)|^{-N} \exp\{-N \operatorname{tr}[\bm{\Gamma}(\Theta)^{-1} \hat{\bm{\Gamma}}]\},
\end{equation}
where $\bm{\Gamma}(\Theta)$ is the array-covariance matrix predicted by Eq.~\eqref{eq:Gamma}. The information gain about the parameters from the sample-covariance matrix is the Kullback--Leibler (KL) divergence of the posterior from the prior,
\begin{equation}
\label{eq:KL}
    \operatorname{D}_\text{KL}[p(\Theta \mid \hat{\bm{\Gamma}}) \parallel p(\Theta)] = \mathbb{E}_{p(\Theta \mid \hat{\bm{\Gamma}})}[\ln p(\Theta \mid \hat{\bm{\Gamma}}) - \ln p(\Theta)].
\end{equation}

There are many numerical methods to calculate the posterior pdf. We use variational inference to find the variational distribution $q(\Theta)$, within a variational family, which minimises the KL divergence from -- and therefore best approximates -- the posterior pdf \citep{Blei_et-al_2017}. Specifically, we construct $q(\Theta)$ from normalising flows \citep{Rezende_Mohamed_2015}. Rearranging Bayes' rule gives
\begin{equation}
    \label{eq:ELBO}
        \ln p(\hat{\bm{\Gamma}}) = \mathcal{L}[q(\Theta)] + \operatorname{D}_\text{KL}[q(\Theta) \parallel p(\Theta \mid \hat{\bm{\Gamma}})],
\end{equation}
where $\mathcal{L}[q(\Theta)] = \mathbb{E}_{q(\Theta)}[\ln p(\Theta, \hat{\bm{\Gamma}}) - \ln q(\Theta)]$ is the evidence lower bound (ELBO). Since the evidence does not depend on $q$, minimising the KL divergence is equivalent to maximising the ELBO. This optimisation -- in our case, of the normalising flow parameters [Sec.~S2, supplementary material] -- is often considerably cheaper than sampling, e.g., with Markov chain Monte Carlo \citep{Zhang_Curtis_2020}. In the examples [Sec.~\ref{sec:examples}], we use a bounded uniform prior [Table~S1, supplementary material]. Posterior statistics on direct and derived properties are computed from an ensemble 20\,000 samples drawn from the variational posterior, and are quoted as the marginal mode with 90\,\% credible-interval bound offsets.

\section{\label{sec:examples} Examples}

\subsection{\label{sec:synthetic}Synthetic abyssal T-phase at hydrophone triplet}

\begin{figure}[bt]
\centerline{\includegraphics[width=6in]{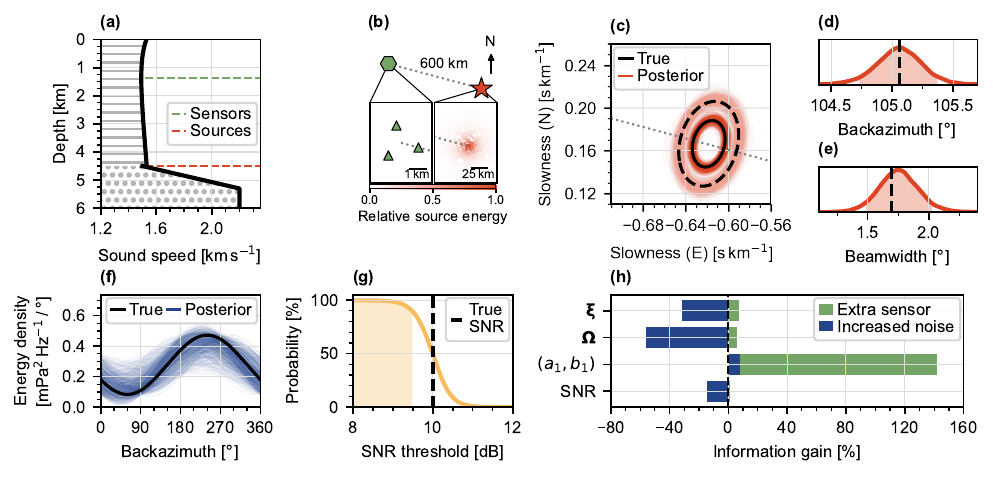}}
\caption{\label{fig:synthetic}Probabilistic wavefield decomposition of a synthetic abyssal T-phase at 5\,Hz. (a) Sound-speed profile (stripes, water; dots, sediment) with source and sensor depths. (b) Source intensity (red) and hydrophone triplet (green; inset), with the backazimuth to the centre of the source patch (dotted). (c) True slowness distribution of signal component (black; solid and dashed contours at 50\,\% and 10\,\% of maximum energy) and posterior samples (red; dark and light contours at 50\,\% and 10\,\% of maximum energy). Marginal posteriors of (d) the backazimuth of the location vector and (e) the backazimuth beamwidth (red; shading, 90\,\% credible interval), with true values (dashed). (f) Angular energy density of the environmental noise component: true (black) and posterior samples (blue). (g) Posterior probability that the signal-to-noise ratio (SNR) exceeds a threshold (yellow; shading, thresholds exceeded with at least 90\,\% probability), with the true SNR (dashed). (h) Change in information gain on the location vector $\bm{\xi}$, scale matrix $\bm{\Omega}$, angular-harmonic noise coefficients $(a_1, b_1)$ and SNR from adding a central hydrophone (green) and from increasing the noise energy by 10\,dB (blue).}
\end{figure}

Abyssal T-phases are generated when seafloor roughness near a deep-ocean earthquake epicentre scatters seismic energy into normal modes of the water column and seafloor sediments \citep{de-Groot-Hedlin_Orcutt_2001}. We synthesise an abyssal T-phase using a \citet{Munk_1974} sound-speed profile [Fig.~\ref{fig:synthetic}(a)] and a patch of incoherently-driven seafloor sources whose strength decreases with distance from the epicentre \citep{de-Groot-Hedlin_Orcutt_2001}, described in detail in Sec.~S3 [supplementary material]. The field is recorded by an IMS-style hydrophone triplet on the axis of the SOFAR channel at a range of 600\,km [Fig.~\ref{fig:synthetic}(a,b)]. As a reference for the inversion, we fit a normal distribution in slowness to the simulated T-phase energy distribution -- the signal component in this example. Dispersion amongst the excited modes gives an apparent-slowness beamwidth of 13\,ms\,km$^{-1}$ about a mean of 640\,ms\,km$^{-1}$, and the spatial extent of the source gives a backazimuth beamwidth of 1.7$^\circ$ about a mean of 105.0$^\circ$ [Fig.~\ref{fig:synthetic}(c)]. We let the self-noise component be negligible, but include an anisotropic angular-harmonic environmental noise component with energy 10\,dB below the signal, propagating at the fundamental-mode slowness of 668\,ms\,km$^{-1}$ [Eq.~\eqref{eq:noise_2D} truncated at $M = 1$, with $a_1 = 30\,\%$ and $b_1 = 18\,\%$; Fig.~\ref{fig:synthetic}(f)].

At each of five frequencies within a quarter-octave band centred on 5\,Hz, two sample-covariance matrices are synthesised: one containing noise only, averaged over 96 realisations, and one containing both the signal and noise, averaged over 16 realisations. The ten matrices are inverted jointly for a normally distributed signal component over 2D slowness [Eq.~\eqref{eq:SN_CF} with $\bm{\alpha} = \bm{0}$] and a 2D angular-harmonic noise component [Eq.~\eqref{eq:noise_CF_2D} truncated at $M = 1$ with fixed $s = 668$\,ms\,km$^{-1}$]. All true signal parameters lie within the 90\,\% posterior credible intervals, which have half-widths about the mode of $\pm 2$\,ms\,km$^{-1}$ for mean apparent slowness, $^{+3}_{-2}$\,ms\,km$^{-1}$ for apparent-slowness beamwidth, and $\pm0.3\,^\circ$ for both mean backazimuth and backazimuth beamwidth [Figs.~\ref{fig:synthetic}(c)--\ref{fig:synthetic}(e)]. The posterior marginal distributions of the angular-harmonic noise coefficients are less well informed, at $a_1 = \left(29_{-31}^{+4}\right)\,\%$ and $b_1 = \left(21_{-28}^{+11}\right)\,\%$, though they peak near the true values [Fig.~\ref{fig:synthetic}(f)]. The signal-to-noise ratio (SNR), defined as the ratio of the total energies of the signal to noise components, is recovered to within $^{+0.6}_{-0.7}$\,dB of its true value of 10\,dB, and the posterior probability that the SNR exceeds 9.5\,dB is 90\,\% [Fig.~\ref{fig:synthetic}(g)].

The information obtained from an inversion depends on both the array design and the wavefield. We quantify this using the information gain \citep{Strutz_Curtis_2024} -- the KL divergence of the posterior from the prior [Eq.~\eqref{eq:KL}], evaluated for the marginal posterior of parameter subsets. Relative to the baseline experiment, we consider two changes: adding a fourth hydrophone at the centre of the triplet, and increasing the environmental noise energy by 10\,dB [Fig.~\ref{fig:synthetic}(h)]. The additional hydrophone increases the information gain by less than 8\,\% for the signal parameters and the SNR; however, it increases by 142\,\% for the angular-harmonic noise coefficients, as the new sensor samples additional inter-sensor azimuths. The increased noise energy reduces the information gain on the SNR, mean and beamwidth of the signal component by 15\,\%, 31\,\% and 56\,\%, respectively.

We test whether the data favour a distributed source ($\mathcal{H}_\text{dist.}$) over a point source ($\mathcal{H}_\text{point}$), the latter modelled with finite apparent-slowness beamwidth but zero backazimuth beamwidth. The posteriors are calculated under each hypothesis, and the evidence $p(\hat{\bm{\Gamma}} \mid \mathcal{H})$ is estimated with the learned harmonic mean estimator \citep{Polanska_et-al_2025}. The ratio of the evidences, the Bayes factor $K$, quantifies the relative support for the two hypotheses. We obtain $\ln K \approx 218$\,nats in favour of $\mathcal{H}_\text{dist.}$, far exceeding the threshold for very strong evidence \citep[$\ln K > 5$\,nats;][]{Kass_Raftery_1995}; this is expected since $\mathcal{H}_\text{dist.}$ is the true model and the information content of the sample-covariance matrices is high.

\subsection{\label{sec:real}Observed thermospheric return at IS37}

\begin{figure}[bt]
\centerline{\includegraphics[width=6in]{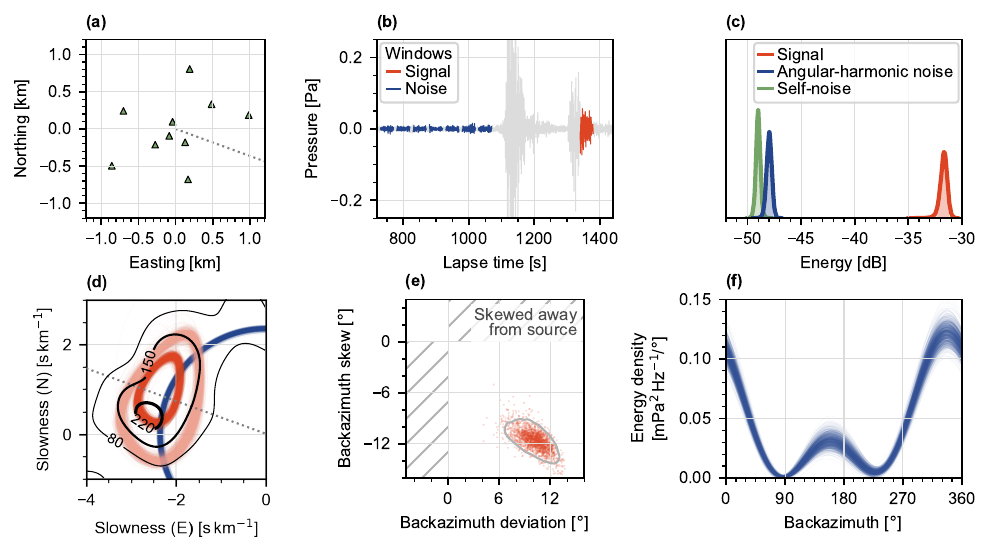}}
\caption{\label{fig:real}Probabilistic wavefield decomposition of the thermospheric return from the Hukkakero explosion recorded at IS37, at 0.5\,Hz. (a) IS37 array sensors (green) and direction to Hukkakero (dotted). (b) Pressure recording against elapsed time from the explosion (grey) with the noise windows (blue) and signal window (red) used in the inversion. (c) Marginal posteriors and 90\,\% credible intervals (shading) of the energies (dB re 1\,Pa$^2$\,Hz$^{-1}$) of the signal (red), angular-harmonic noise (blue) and sensor self-noise (green). (d) Posterior samples of the signal slowness distribution (red; dark and light contours at 50\,\% and 10\,\% of maximum energy) and angular-harmonic noise (blue), with beam power of the signal window (black contours, mPa$^{2}$\,Hz$^{-1}$) and the direction to Hukkakero (dotted). (e) Posterior samples (red) of backazimuth deviation of the location vector from the direction to Hukkakero and the backazimuth skew, with 90\,\% credible region (grey); hatching marks skew away from the source. (f) Directional energy density of the posterior angular-harmonic noise samples.}
\end{figure}

We analyse the slowness distribution of infrasound energy recorded at IS37, a microbarometer array near Bardufoss, Norway [Fig.~\ref{fig:real}(a)], generated by a surface chemical explosion of approximately 20\,t at 11:29:59\,UTC on 19 August 2016. The explosion was one of the annual ammunition-disposal detonations at Hukkakero, northern Finland \citep{Gibbons_et-al_2007}, approximately 321\,km from IS37 at a backazimuth of 110$^\circ$. The recording contains two distinct arrivals [Fig.~\ref{fig:real}(b)]: the first with a celerity typical of stratospheric returns, and the second typical of thermospheric returns \citep{Nippress_et-al_2014}.

We decompose the thermospheric return into three wavefield components: a skew-normal signal component [Eq.~\eqref{eq:SN_CF}], an angular-harmonic noise component [Eq.~\eqref{eq:noise_CF_2D}, truncated at $M = 2$] and a sensor self-noise component. Two sample-covariance matrices are estimated at each of 0.4, 0.5 and 0.6\,Hz: one from five consecutive noise windows immediately preceding the stratospheric arrival, and the other from a single signal window [Fig.~\ref{fig:real}(b)]. Within each 40\,s-long window, cross-spectral densities are estimated with 10 Slepian tapers, giving $N = 50$ realisations for the noise matrices and $N = 10$ for the signal matrices. All six matrices are inverted jointly: the noise components are shared between the signal and noise windows, whereas the signal component contributes only to the signal window.

The SNR is $(13.8^{+0.5}_{-0.7})$\,dB [Fig.~\ref{fig:real}(c)]; the angular-harmonic noise energy is $\left(1.0 \pm 0.7\right)$\,dB larger than the self-noise. The signal has an apparent slowness of $(2.88^{+0.06}_{-0.08})$\,s\,km$^{-1}$ with a beamwidth of $(0.41 \pm 0.04)$\,s\,km$^{-1}$, and its backazimuth is deflected $(10^{+2}_{-3})^\circ$ clockwise from the direction of Hukkakero, with a beamwidth of $(13 \pm 1)^\circ$ [Fig.~\ref{fig:real}(d)]. In relation to the deflection of the location, the distribution is skewed towards lower apparent slowness by $(0.60 \pm 0.08)$\,s\,km$^{-1}$ and anticlockwise in backazimuth, back towards the source, by $(11 \pm 2)^\circ$. Over 99\,\% of the ensemble [Fig.~\ref{fig:real}(e)] supports the hypothesis that, while cross-winds deflect the directional energy distribution, its shape is skewed back towards the source direction \citep{Smith_et-al_2026}; this causes an additional cross-phase which deviates the direction of the maximum beam power down-skew from the direction of maximum energy [Eq.~\eqref{eq:SN_CF}; Sec.~S1, supplementary material]. The angular-harmonic noise has an apparent slowness of $(2.36 \pm 0.05)$\,s\,km$^{-1}$ and is strongly anisotropic, with two energy peaks: one at a backazimuth of $(158^{+3}_{-4})^\circ$, towards the northern Gulf of Bothnia, and one at $(336 \pm 3)^\circ$, towards the Norwegian Sea, the latter $3.7^{+1.0}_{-0.7}$ times more energetic [Fig.~\ref{fig:real}(f)].

\section{Discussion and conclusions}

The complex Wishart likelihood [Eq.~\eqref{eq:likelihood}] rests on assumptions that are rarely satisfied exactly, and most violations make the posterior over-confident rather than biased. Non-Gaussian wavefields leave the sample covariance unbiased but increase its variance beyond that of the Wishart distribution \citep{Tough_et-al_1995}. The effective number of independent realisations is smaller than $N$ when windows overlap \citep{Carter_et-al_1973} or the spectrum varies across the multitaper bandwidth \citep{Thomson_1982}; replacing $N$ with an effective number of degrees of freedom mitigates both \citep{Bortel_Sovka_2007}, and could also be used to factor correlation between nearby frequency bins. Differences between sensors not represented in the forward model, such as timing offsets, gain mismatches, reversed polarity and unequal self-noise, instead bias the posterior \citep{Bishop_et-al_2020}, so data quality must be verified before inversion. Similarly, treating a quasi-planar array as flat causes systematic errors in the inter-sensor cross-phase \citep{Nouvellet_et-al_2014}; these can be accounted for by a 3D inversion, with an informative prior on the propagation velocity since the vertical slowness is poorly constrained by quasi-planar data.

The forward problem is restricted here to incoherently distributed wavefields, for which cross-terms between distinct slowness components vanish in expectation [Eq.~\eqref{eq:Gamma}]. This is one end member of a spectrum that extends through partially coherent to coherently distributed wavefields \citep{Valaee_et-al_1995}. Mutual incoherence is often an adequate approximation in the far field when the coherence length of the source is short relative to its extent \citep{Carter_Wolf_1977}, although \citet{Smith_et-al_2026} suggest that the scattered Hukkakero return retains observable partial coherence.

Coherence loss between sensors carries information about the wavefield. Parameterising its energy distribution over slowness yields estimates with quantified uncertainties that can feed directly into location and monitoring decisions, and, as shown in the examples, support experimental design, model comparison and hypothesis testing. Natural extensions of the forward model include: partially coherent wavefields, which would allow the retained-coherence hypothesis to be tested; and multicomponent sensors recording vector wavefields, which would allow wave types to be separated by the polarisation of their particle motion as well as their slowness.

\section*{Supplementary material}

See supplementary material for a detailed description of (S1) the effect of skew on beam power, (S2) variational inference by normalising flows, and (S3) the abyssal T-phase simulation.

\section*{Acknowledgements}

J.-A.S. and C.E.-P. are supported by NERC Edinburgh Earth Ecology and Environment Doctoral Training Partnership grant NE/S007407/1. IS37 data were made available by the UK National Data Centre at AWE during an industrial placement undertaken by J.-A.S. \citep{Smith_et-al_2026}. We thank NORSAR for operating the station and Xuebin~Zhao for guidance on using the Python package GeoPVI for variational inference \citep{Zhao_Curtis_2026a}. J.-A.S. thanks David~N.~Green for discussions at AWE~Blacknest that motivated this work.

\section*{Author declarations}

The authours have no conflicts of interest to disclose.

\section*{Data availability}

The open-source Python package Topophone2 provides tools to perform probabilistic wavefield decomposition and analyse the resulting posteriors \citep{Smith_Elijas-Parra_2026}. The synthetic hydroacoustic data were computed with the normal-mode program KRAKEN from the Acoustics Toolbox \citep{Porter_1992}, using the parameters given in Sec.~S3 [supplementary material]. The IS37 infrasound station is part of the International Monitoring System of the Preparatory Commission for the Comprehensive Nuclear-Test-Ban Treaty Organization, whereby data access can be granted via the virtual Data Exploitation Centre of the International Data Centre at \url{https://www.ctbto.org/specials/vdec/}.

%=======================================================

%\nocite{Azzalini_Capitanio_1999}
%\bibliographystyle{jasaauthyear2}
%\bibliography{bibliography}

\begin{thebibliography}{44}
\def\enquote#1{``#1,''}
\def\plainquote#1{``#1''}
\expandafter\ifx\csname natexlab\endcsname\relax\def\natexlab#1{#1}\fi
\providecommand{\dourl}[1]{\href{http://#1}{\nolinkurl{#1}}}
\providecommand{\bibinfo}[2]{#2}
\providecommand{\noopsort}[1]{}
\providecommand{\switchargs}[2]{#2#1}
  \def\eatspace #1{#1}

\bibitem[{Anderson \emph{et~al.}(2007)Anderson, Fagan, Tinker, Kraft, and Hutchenson}]{Anderson_et-al_2007}
\bibinfo{author}{Anderson, D.~N.}, \bibinfo{author}{Fagan, D.~K.}, \bibinfo{author}{Tinker, M.~A.}, \bibinfo{author}{Kraft, G.~D.},  and \bibinfo{author}{Hutchenson, K.~D.} (\textbf{\bibinfo{year}{2007}}). \enquote{\bibinfo{title}{{A Mathematical Statistics Formulation of the Teleseismic Explosion Identification Problem with Multiple Discriminants}}} \bibinfo{journal}{Bulletin of the Seismological Society of America} \textbf{97}(5), \bibinfo{pages}{1730--1741}, \dodoi{10.1785/0120060052}.

\bibitem[{Bartlett(1948)}]{Bartlett_1948}
\bibinfo{author}{Bartlett, M.~S.} (\textbf{\bibinfo{year}{1948}}). \enquote{\bibinfo{title}{{Smoothing Periodograms from Time-Series with Continuous Spectra}}} \bibinfo{journal}{Nature} \textbf{161}, \bibinfo{pages}{686--687}, \dodoi{10.1038/161686a0}.

\bibitem[{Bishop \emph{et~al.}(2020)Bishop, Fee, and Szuberla}]{Bishop_et-al_2020}
\bibinfo{author}{Bishop, J.~W.}, \bibinfo{author}{Fee, D.},  and \bibinfo{author}{Szuberla, C. A.~L.} (\textbf{\bibinfo{year}{2020}}). \enquote{\bibinfo{title}{{Improved infrasound array processing with robust estimators}}} \bibinfo{journal}{Geophysical Journal International} \textbf{221}(3), \bibinfo{pages}{2058--2074}, \dodoi{10.1093/gji/ggaa110}.

\bibitem[{Blandford(1974)}]{Blandford_1974}
\bibinfo{author}{Blandford, R.~R.} (\textbf{\bibinfo{year}{1974}}). \enquote{\bibinfo{title}{{An automatic event detector at the Tonto Forest Seismic Observatory}}} \bibinfo{journal}{Geophysics} \textbf{39}(5), \bibinfo{pages}{633--643}, \dodoi{10.1190/1.1440453}.

\bibitem[{Blei \emph{et~al.}(2017)Blei, Kucukelbir, and McAuliffe}]{Blei_et-al_2017}
\bibinfo{author}{Blei, D.~M.}, \bibinfo{author}{Kucukelbir, A.},  and \bibinfo{author}{McAuliffe, J.~D.} (\textbf{\bibinfo{year}{2017}}). \enquote{\bibinfo{title}{{Variational Inference: A Review for Statisticians}}} \bibinfo{journal}{Journal of the American Statistical Association} \textbf{112}(518), \bibinfo{pages}{859--877}, \dodoi{10.1080/01621459.2017.1285773}.

\bibitem[{Bortel and Sovka(2007)}]{Bortel_Sovka_2007}
\bibinfo{author}{Bortel, R.},  and \bibinfo{author}{Sovka, P.} (\textbf{\bibinfo{year}{2007}}). \enquote{\bibinfo{title}{{Approximation of statistical distribution of magnitude squared coherence estimated with segment overlapping}}} \bibinfo{journal}{Signal Processing} \textbf{87}(5), \bibinfo{pages}{1100--1117}, \dodoi{10.1016/j.sigpro.2006.10.003}.

\bibitem[{Capon(1969)}]{Capon_1969}
\bibinfo{author}{Capon, J.} (\textbf{\bibinfo{year}{1969}}). \enquote{\bibinfo{title}{{High-resolution frequency-wavenumber spectrum analysis}}} \bibinfo{journal}{Proceedings of the IEEE} \textbf{57}(8), \bibinfo{pages}{1408--1418}, \dodoi{10.1109/PROC.1969.7278}.

\bibitem[{Carter \emph{et~al.}(1973)Carter, Knapp, and Nuttall}]{Carter_et-al_1973}
\bibinfo{author}{Carter, G.}, \bibinfo{author}{Knapp, C.},  and \bibinfo{author}{Nuttall, A.} (\textbf{\bibinfo{year}{1973}}). \enquote{\bibinfo{title}{{Estimation of the magnitude-squared coherence function via overlapped fast Fourier transform processing}}} \bibinfo{journal}{IEEE Transactions on Audio and Electroacoustics} \textbf{21}(4), \bibinfo{pages}{337--344}, \dodoi{10.1109/TAU.1973.1162496}.

\bibitem[{Carter and Wolf(1977)}]{Carter_Wolf_1977}
\bibinfo{author}{Carter, W.~H.},  and \bibinfo{author}{Wolf, E.} (\textbf{\bibinfo{year}{1977}}). \enquote{\bibinfo{title}{{Coherence and radiometry with quasihomogeneous planar sources}}} \bibinfo{journal}{Journal of the Optical Society of America} \textbf{67}(6), \bibinfo{pages}{785--796}, \dodoi{10.1364/JOSA.67.000785}.

\bibitem[{Cox(1973)}]{Cox_1973}
\bibinfo{author}{Cox, H.} (\textbf{\bibinfo{year}{1973}}). \enquote{\bibinfo{title}{{Spatial correlation in arbitrary noise fields with application to ambient sea noise}}} \bibinfo{journal}{The Journal of the Acoustical Society of America} \textbf{54}(5), \bibinfo{pages}{1289--1301}, \dodoi{10.1121/1.1914426}.

\bibitem[{de~Groot-Hedlin and Orcutt(2001)}]{de-Groot-Hedlin_Orcutt_2001}
\bibinfo{author}{de~Groot-Hedlin, C.~D.},  and \bibinfo{author}{Orcutt, J.~A.} (\textbf{\bibinfo{year}{2001}}). \enquote{\bibinfo{title}{{Excitation of T-phases by seafloor scattering}}} \bibinfo{journal}{The Journal of the Acoustical Society of America} \textbf{109}(5), \bibinfo{pages}{1944--1954}, \dodoi{10.1121/1.1361057}.

\bibitem[{Galetti \emph{et~al.}(2015)Galetti, Curtis, Meles, and Baptie}]{Galetti_et-al_2015}
\bibinfo{author}{Galetti, E.}, \bibinfo{author}{Curtis, A.}, \bibinfo{author}{Meles, G.~A.},  and \bibinfo{author}{Baptie, B.} (\textbf{\bibinfo{year}{2015}}). \enquote{\bibinfo{title}{{Uncertainty Loops in Travel-Time Tomography from Nonlinear Wave Physics}}} \bibinfo{journal}{Phys. Rev. Lett.} \textbf{114}(14), \bibinfo{pages}{148501}, \dodoi{10.1103/PhysRevLett.114.148501}.

\bibitem[{Gibbons \emph{et~al.}(2007)Gibbons, Ringdal, and Kv{\ae}rna}]{Gibbons_et-al_2007}
\bibinfo{author}{Gibbons, S.~J.}, \bibinfo{author}{Ringdal, F.},  and \bibinfo{author}{Kv{\ae}rna, T.} (\textbf{\bibinfo{year}{2007}}). \enquote{\bibinfo{title}{{Joint seismic-infrasonic processing of recordings from a repeating source of atmospheric explosions}}} \bibinfo{journal}{The Journal of the Acoustical Society of America} \textbf{122}(5), \bibinfo{pages}{EL158--EL164}, \dodoi{10.1121/1.2784533}.

\bibitem[{Gibbons \emph{et~al.}(2008)Gibbons, Ringdal, and Kv{\ae}rna}]{Gibbons_et-al_2008}
\bibinfo{author}{Gibbons, S.~J.}, \bibinfo{author}{Ringdal, F.},  and \bibinfo{author}{Kv{\ae}rna, T.} (\textbf{\bibinfo{year}{2008}}). \enquote{\bibinfo{title}{{Detection and characterization of seismic phases using continuous spectral estimation on incoherent and partially coherent arrays}}} \bibinfo{journal}{Geophysical Journal International} \textbf{172}(1), \bibinfo{pages}{405--421}, \dodoi{10.1111/j.1365-246X.2007.03650.x}.

\bibitem[{Goodman(1963)}]{Goodman_1963}
\bibinfo{author}{Goodman, N.~R.} (\textbf{\bibinfo{year}{1963}}). \enquote{\bibinfo{title}{{Statistical Analysis Based on a Certain Multivariate Complex Gaussian Distribution (An Introduction)}}} \bibinfo{journal}{The Annals of Mathematical Statistics} \textbf{34}(1), \bibinfo{pages}{152--177}, \dodoi{10.1214/aoms/1177704250}.

\bibitem[{Gossard(1969)}]{Gossard_1969}
\bibinfo{author}{Gossard, E.~E.} (\textbf{\bibinfo{year}{1969}}). \enquote{\bibinfo{title}{{The effect of bandwidth on the interpretation of the cross spectra of wave recordings from spatially separated sites}}} \bibinfo{journal}{Journal of Geophysical Research (1896-1977)} \textbf{74}(1), \bibinfo{pages}{325--337}, \dodoi{10.1029/JA074i001p00325}.

\bibitem[{Green(2015)}]{Green_2015}
\bibinfo{author}{Green, D.~N.} (\textbf{\bibinfo{year}{2015}}). \enquote{\bibinfo{title}{{The spatial coherence structure of infrasonic waves: analysis of data from International Monitoring System arrays}}} \bibinfo{journal}{Geophysical Journal International} \textbf{201}(1), \bibinfo{pages}{377--389}, \dodoi{10.1093/gji/ggu495}.

\bibitem[{Kass and Raftery(1995)}]{Kass_Raftery_1995}
\bibinfo{author}{Kass, R.~E.},  and \bibinfo{author}{Raftery, A.~E.} (\textbf{\bibinfo{year}{1995}}). \enquote{\bibinfo{title}{{Bayes Factors}}} \bibinfo{journal}{Journal of the American Statistical Association} \textbf{90}(430), \bibinfo{pages}{773--795}, \dodoi{10.1080/01621459.1995.10476572}.

\bibitem[{Kim and Genton(2011)}]{Kim_Genton_2011}
\bibinfo{author}{Kim, H.-M.},  and \bibinfo{author}{Genton, M.~G.} (\textbf{\bibinfo{year}{2011}}). \enquote{\bibinfo{title}{{Characteristic functions of scale mixtures of multivariate skew-normal distributions}}} \bibinfo{journal}{Journal of Multivariate Analysis} \textbf{102}(7), \bibinfo{pages}{1105--1117}, \dodoi{10.1016/j.jmva.2011.03.004}.

\bibitem[{Mack and Flinn(1971)}]{Mack_Flinn_1971}
\bibinfo{author}{Mack, H.},  and \bibinfo{author}{Flinn, E.~A.} (\textbf{\bibinfo{year}{1971}}). \enquote{\bibinfo{title}{{Analysis of the Spatial Coherence of Short-Period Acoustic-Gravity Waves in the Atmosphere}}} \bibinfo{journal}{Geophysical Journal International} \textbf{26}(1--4), \bibinfo{pages}{255--269}, \dodoi{10.1111/j.1365-246X.1971.tb03399.x}.

\bibitem[{Modrak \emph{et~al.}(2010)Modrak, Arrowsmith, and Anderson}]{Modrak_et-al_2010}
\bibinfo{author}{Modrak, R.~T.}, \bibinfo{author}{Arrowsmith, S.~J.},  and \bibinfo{author}{Anderson, D.~N.} (\textbf{\bibinfo{year}{2010}}). \enquote{\bibinfo{title}{{A Bayesian framework for infrasound location}}} \bibinfo{journal}{Geophysical Journal International} \textbf{181}(1), \bibinfo{pages}{399--405}, \dodoi{10.1111/j.1365-246X.2010.04499.x}.

\bibitem[{Munk(1974)}]{Munk_1974}
\bibinfo{author}{Munk, W.~H.} (\textbf{\bibinfo{year}{1974}}). \enquote{\bibinfo{title}{{Sound channel in an exponentially stratified ocean, with application to SOFAR}}} \bibinfo{journal}{The Journal of the Acoustical Society of America} \textbf{55}(2), \bibinfo{pages}{220--226}, \dodoi{10.1121/1.1914492}.

\bibitem[{Nippress \emph{et~al.}(2014)Nippress, Green, Marcillo, and Arrowsmith}]{Nippress_et-al_2014}
\bibinfo{author}{Nippress, A.}, \bibinfo{author}{Green, D.~N.}, \bibinfo{author}{Marcillo, O.~E.},  and \bibinfo{author}{Arrowsmith, S.~J.} (\textbf{\bibinfo{year}{2014}}). \enquote{\bibinfo{title}{{Generating regional infrasound celerity-range models using ground-truth information and the implications for event location}}} \bibinfo{journal}{Geophysical Journal International} \textbf{197}(2), \bibinfo{pages}{1154--1165}, \dodoi{10.1093/gji/ggu049}.

\bibitem[{Nouvellet \emph{et~al.}(2014)Nouvellet, Charbit, Roueff, and Le~Pichon}]{Nouvellet_et-al_2014}
\bibinfo{author}{Nouvellet, A.}, \bibinfo{author}{Charbit, M.}, \bibinfo{author}{Roueff, F.},  and \bibinfo{author}{Le~Pichon, A.} (\textbf{\bibinfo{year}{2014}}). \enquote{\bibinfo{title}{{Slowness estimation from noisy time delays observed on non-planar arrays}}} \bibinfo{journal}{Geophysical Journal International} \textbf{198}(2), \bibinfo{pages}{1199--1207}, \dodoi{10.1093/gji/ggu197}.

\bibitem[{Oberhettinger(1973)}]{Oberhettinger_1973}
\bibinfo{author}{Oberhettinger, F.} (\textbf{\bibinfo{year}{1973}}). {Probability and Mathematical Statistics} \emph{\bibinfo{title}{{Fourier Transforms of Distributions and their Inverses}}} (\bibinfo{publisher}{Academic Press}).

\bibitem[{Polanska \emph{et~al.}(2025)Polanska, Price, Piras, Spurio~Mancini, and McEwen}]{Polanska_et-al_2025}
\bibinfo{author}{Polanska, A.}, \bibinfo{author}{Price, M.~A.}, \bibinfo{author}{Piras, D.}, \bibinfo{author}{Spurio~Mancini, A.},  and \bibinfo{author}{McEwen, J.~D.} (\textbf{\bibinfo{year}{2025}}). \enquote{\bibinfo{title}{{Learned harmonic mean estimation of the Bayesian evidence with normalizing flows}}} \bibinfo{journal}{The Open Journal of Astrophysics} \textbf{8}, \dodoi{10.33232/001c.146026}.

\bibitem[{Porter(1992)}]{Porter_1992}
\bibinfo{author}{Porter, M.~B.} (\textbf{\bibinfo{year}{1992}}). \emph{\bibinfo{title}{{The KRAKEN Normal Mode Program}}}, \bibinfo{organization}{U.S. Department of Commerce, National Technical Information Service}.

\bibitem[{Poste \emph{et~al.}(2023)Poste, Charbit, Le~Pichon, Listowski, Roueff, and Vergoz}]{Poste_et-al_2023}
\bibinfo{author}{Poste, B.}, \bibinfo{author}{Charbit, M.}, \bibinfo{author}{Le~Pichon, A.}, \bibinfo{author}{Listowski, C.}, \bibinfo{author}{Roueff, F.},  and \bibinfo{author}{Vergoz, J.} (\textbf{\bibinfo{year}{2023}}). \enquote{\bibinfo{title}{{The multichannel maximum-likelihood (MCML) method: a new approach for infrasound detection and wave parameter estimation}}} \bibinfo{journal}{Geophysical Journal International} \textbf{232}(2), \bibinfo{pages}{1099--1112}, \dodoi{10.1093/gji/ggac377}.

\bibitem[{Rezende and Mohamed(2015)}]{Rezende_Mohamed_2015}
\bibinfo{author}{Rezende, D.},  and \bibinfo{author}{Mohamed, S.} (\textbf{\bibinfo{year}{2015}}). \enquote{\bibinfo{title}{{Variational Inference with Normalizing Flows}}} in \emph{\bibinfo{booktitle}{{Proceedings of the 32nd International Conference on Machine Learning}}}, edited by \bibinfo{editor}{F.~Bach} and \bibinfo{editor}{D.~Blei}, \bibinfo{publisher}{PMLR}, pp. \bibinfo{pages}{1530--1538}.

\bibitem[{Selby(2008)}]{Selby_2008}
\bibinfo{author}{Selby, N.~D.} (\textbf{\bibinfo{year}{2008}}). \enquote{\bibinfo{title}{{Application of a Generalized F Detector at a Seismometer Array}}} \bibinfo{journal}{Bulletin of the Seismological Society of America} \textbf{98}(5), \bibinfo{pages}{2469--2481}, \dodoi{10.1785/0120070282}.

\bibitem[{Smith and Elijas-Parra(2026)}]{Smith_Elijas-Parra_2026}
\bibinfo{author}{Smith, J.-A.},  and \bibinfo{author}{Elijas-Parra, C.} (\textbf{\bibinfo{year}{2026}}). \plainquote{\bibinfo{title}{{Topophone2: Probabilistic wavefield decomposition from array covariance}}} \bibinfo{howpublished}{Zenodo}, \dodoi{10.5281/zenodo.23731332}, \bibinfo{note}{version 1.0.0, computer software}.

\bibitem[{Smith \emph{et~al.}(2026)Smith, Green, and Nippress}]{Smith_et-al_2026}
\bibinfo{author}{Smith, J.-A.}, \bibinfo{author}{Green, D.~N.},  and \bibinfo{author}{Nippress, S. E.~J.} (\textbf{\bibinfo{year}{2026}}). \enquote{\bibinfo{title}{{Characterisation of scattered transient infrasonic wavefields combining array and single-sensor measures}}} \bibinfo{journal}{The Journal of the Acoustical Society of America} \textbf{160}(1), \bibinfo{pages}{148--159}, \dodoi{10.1121/10.0044268}.

\bibitem[{Sober(2015)}]{Sober_2015}
\bibinfo{author}{Sober, E.} (\textbf{\bibinfo{year}{2015}}). \enquote{\bibinfo{title}{{2 -- The probabilistic turn}}} in \emph{\bibinfo{booktitle}{{Ockham’s Razors: A User’s Manual}}} (\bibinfo{publisher}{{Cambridge University Press}}), pp. \bibinfo{pages}{61--152}, \dodoi{10.1017/CBO9781107705937}.

\bibitem[{Strutz and Curtis(2024)}]{Strutz_Curtis_2024}
\bibinfo{author}{Strutz, D.},  and \bibinfo{author}{Curtis, A.} (\textbf{\bibinfo{year}{2024}}). \enquote{\bibinfo{title}{{Variational Bayesian experimental design for geophysical applications: seismic source location, amplitude versus offset inversion, and estimating CO2 saturations in a subsurface reservoir}}} \bibinfo{journal}{Geophysical Journal International} \textbf{236}(3), \bibinfo{pages}{1309--1331}, \dodoi{10.1093/gji/ggad492}.

\bibitem[{Szuberla and Olson(2003)}]{Szuberla_Olson_2003}
\bibinfo{author}{Szuberla, C. A.~L.},  and \bibinfo{author}{Olson, J.~V.} (\textbf{\bibinfo{year}{2003}}). \enquote{\bibinfo{title}{{Uncertainties associated with parameter estimation in atmospheric infrasound arrays}}} \bibinfo{journal}{The Journal of the Acoustical Society of America} \textbf{115}(1), \bibinfo{pages}{253--258}, \dodoi{10.1121/1.1635407}.

\bibitem[{Thomson(1982)}]{Thomson_1982}
\bibinfo{author}{Thomson, D.~J.} (\textbf{\bibinfo{year}{1982}}). \enquote{\bibinfo{title}{{Spectrum estimation and harmonic analysis}}} \bibinfo{journal}{Proceedings of the IEEE} \textbf{70}(9), \bibinfo{pages}{1055--1096}, \dodoi{10.1109/PROC.1982.12433}.

\bibitem[{Tough \emph{et~al.}(1995)Tough, Blacknell, and Quegan}]{Tough_et-al_1995}
\bibinfo{author}{Tough, R. J.~A.}, \bibinfo{author}{Blacknell, D.},  and \bibinfo{author}{Quegan, S.} (\textbf{\bibinfo{year}{1995}}). \enquote{\bibinfo{title}{{A Statistical Description of Polarimetric and Interferometric Synthetic Aperture Radar Data}}} \bibinfo{journal}{Proceedings of the Royal Society of London. Series A: Mathematical and Physical Sciences} \textbf{449}(1937), \bibinfo{pages}{567--589}, \dodoi{10.1098/rspa.1995.0059}.

\bibitem[{Trump and Ottersten(1996)}]{Trump_Ottersten_1996}
\bibinfo{author}{Trump, T.},  and \bibinfo{author}{Ottersten, B.} (\textbf{\bibinfo{year}{1996}}). \enquote{\bibinfo{title}{{Estimation of nominal direction of arrival and angular spread using an array of sensors}}} \bibinfo{journal}{Signal Processing} \textbf{50}(1), \bibinfo{pages}{57--69}, \dodoi{10.1016/0165-1684(96)00003-5}.

\bibitem[{Valaee \emph{et~al.}(1995)Valaee, Champagne, and Kabal}]{Valaee_et-al_1995}
\bibinfo{author}{Valaee, S.}, \bibinfo{author}{Champagne, B.},  and \bibinfo{author}{Kabal, P.} (\textbf{\bibinfo{year}{1995}}). \enquote{\bibinfo{title}{{Parametric localization of distributed sources}}} \bibinfo{journal}{IEEE Transactions on Signal Processing} \textbf{43}(9), \bibinfo{pages}{2144--2153}, \dodoi{10.1109/78.414777}.

\bibitem[{Williams \emph{et~al.}(2026)Williams, Picard, and Anderson}]{Williams_et-al_2026}
\bibinfo{author}{Williams, B.~J.}, \bibinfo{author}{Picard, R.},  and \bibinfo{author}{Anderson, D.~N.} (\textbf{\bibinfo{year}{2026}}). \enquote{\bibinfo{title}{{Uncertainty estimation for multiphenomenology explosion monitoring}}} \bibinfo{journal}{Geophysical Journal International} \textbf{245}(3), \bibinfo{pages}{ggag125}, \dodoi{10.1093/gji/ggag125}.

\bibitem[{Wu \emph{et~al.}(2019)Wu, Deng, Li, Li, and Huang}]{Wu_et-al_2019}
\bibinfo{author}{Wu, T.}, \bibinfo{author}{Deng, Z.}, \bibinfo{author}{Li, Y.}, \bibinfo{author}{Li, Z.},  and \bibinfo{author}{Huang, Y.} (\textbf{\bibinfo{year}{2019}}). \enquote{\bibinfo{title}{{Estimation of Two-Dimensional Non-Symmetric Incoherently Distributed Source with L-Shape Arrays}}} \bibinfo{journal}{Sensors} \textbf{19}(5), \bibinfo{pages}{1226}, \dodoi{10.3390/s19051226}.

\bibitem[{Xenaki \emph{et~al.}(2016)Xenaki, Fernandez-Grande, and Gerstoft}]{Xenaki_et-al_2016}
\bibinfo{author}{Xenaki, A.}, \bibinfo{author}{Fernandez-Grande, E.},  and \bibinfo{author}{Gerstoft, P.} (\textbf{\bibinfo{year}{2016}}). \enquote{\bibinfo{title}{{Block-sparse beamforming for spatially extended sources in a Bayesian formulation}}} \bibinfo{journal}{The Journal of the Acoustical Society of America} \textbf{140}(3), \bibinfo{pages}{1828--1838}, \dodoi{10.1121/1.4962325}.

\bibitem[{Zhang and Curtis(2020)}]{Zhang_Curtis_2020}
\bibinfo{author}{Zhang, X.},  and \bibinfo{author}{Curtis, A.} (\textbf{\bibinfo{year}{2020}}). \enquote{\bibinfo{title}{{Seismic Tomography Using Variational Inference Methods}}} \bibinfo{journal}{Journal of Geophysical Research: Solid Earth} \textbf{125}(4), \bibinfo{pages}{e2019JB018589}, \dodoi{10.1029/2019JB018589}.

\bibitem[{Zhao and Curtis(2026)}]{Zhao_Curtis_2026a}
\bibinfo{author}{Zhao, X.},  and \bibinfo{author}{Curtis, A.} (\textbf{\bibinfo{year}{2026}}). \enquote{\bibinfo{title}{{GeoPVI: a Python package for geoscientific inversion using parametric variational inference}}} \bibinfo{journal}{Geophysical Journal International} \textbf{245}(3), \bibinfo{pages}{ggag138}, \dodoi{10.1093/gji/ggag138}.

\end{thebibliography}

%=======================================================

\clearpage
\includepdf[pages=-]{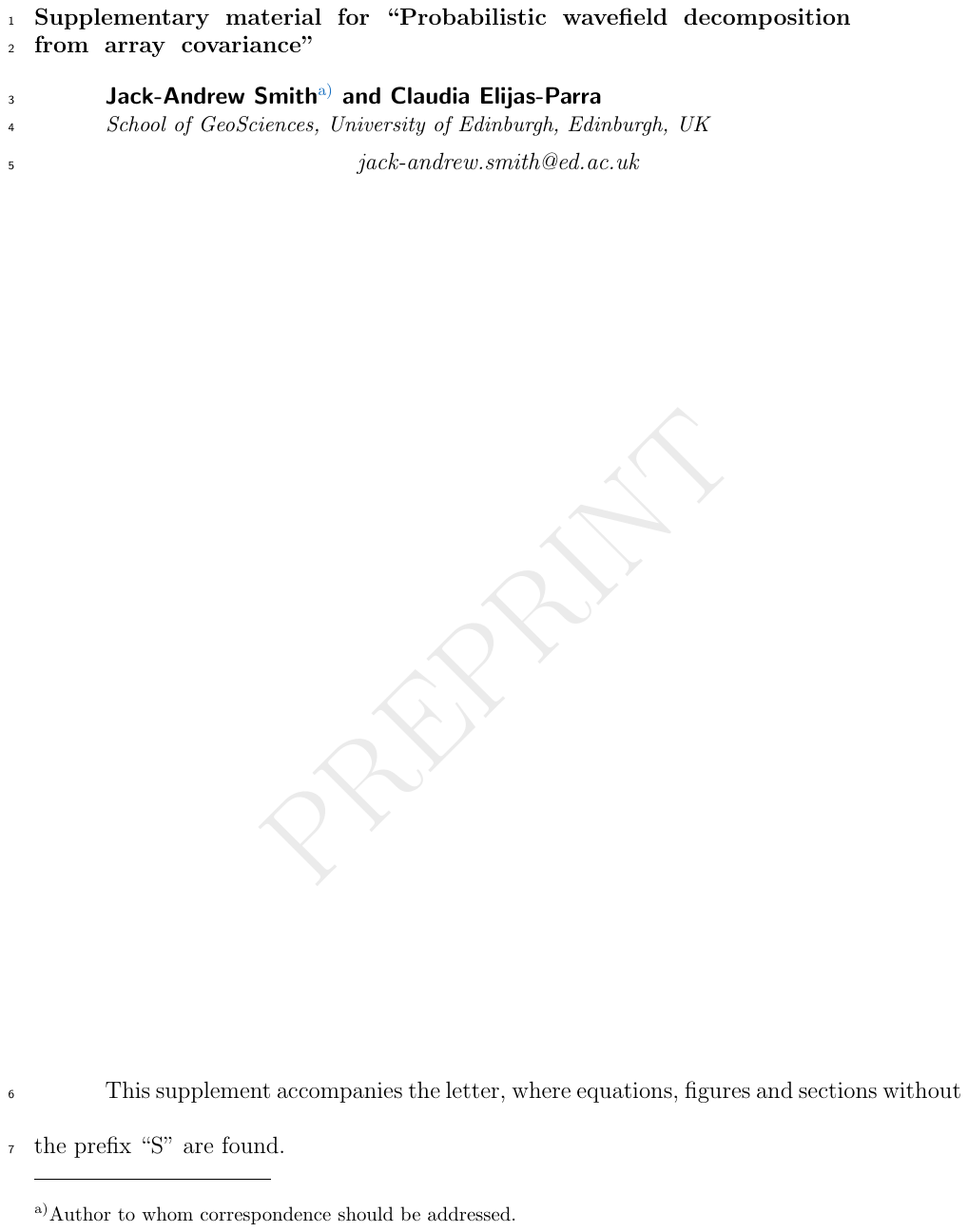}

\end{document}